\documentclass[preprint,pra,amsmath,amssymb]{revtex4}
\usepackage{graphicx}
\usepackage{dcolumn}
\usepackage{epic}%
\usepackage{eepic}%
\usepackage{bm}

\begin{document}

\preprint{APS/123-QED}

\newcommand{\zlabel}{\label} 
\newcommand{\hH}{\hat{H}} 
\newcommand{\hT}{\hat{T}} 
\newcommand{\ke}[1]{-\mbox{$\frac{1}{2}$}\nabla_{{#1}}^2} 
\newcommand{\roo}{{\rho_1}} 
\newcommand{\hroo}{{\hat{\rho}_1}} 
\newcommand{\hvroo}{{\hat{\varrho}_1}} 
\newcommand{\vroo}{{\varrho}_1} 
\newcommand{\kp}{\kappa} 
\newcommand{\hkp}{\hat{\kappa}} 
\newcommand{\rc}[1]{r^{-1}_{#1}} 
\newcommand{\fc}{\frac} 
\newcommand{\lt}{\left} 
\newcommand{\rt}{\right} 
\newcommand{\ran}{\rangle} 
\newcommand{\lan}{\langle} 
\newcommand{\cE}{{\cal E}} 
\newcommand{\hVee}{\hat{V}_{\text{ee}}} 
\newcommand{\hV}{\hat{V}} 
\newcommand{\hGm}{\hat{\Gamma}}
\newcommand{\Gm}{\Gamma}
\newcommand{\gm}{\gamma}
\newcommand{\si}{\sigma}
\newcommand{\dg}{\dagger} 
\newcommand{\mr}{\mathbf{r}}
\newcommand{\mx}{\mathbf{x}}
\newcommand{\om}{\omega}
\newcommand{\hE}{\hat{E}}
\newcommand{\dt}{\delta}
\newcommand{\al}{\alpha} 
\newcommand{\be}{\beta} 
\newcommand{\hs}[1]{\hspace{#1ex}}
\newcommand{\vs}[1]{\vspace{#1ex}}
\newcommand{\pr}{\prime}
\newcommand{\hvj}{\hat{v}_{J}} 
\newcommand{\hvxc}{\hat{v}_{\text{xc}}} 
\newcommand{\vxc}{v_{\text{xc}}} 
\newcommand{\hvx}{\hat{v}_{\text{x}}} 
\newcommand{\hvco}{\hat{v}_{\text{co}}} 
\newcommand{\vco}{v_{\text{co}}} 
\newcommand{\hvcoroo}{\hat{v}_{\text{co}}^{\rho_1}} 
\newcommand{\vcoroo}{v_{\text{co}}^{\rho_1}} 
\newcommand{\hvxroo}{\hat{v}_{\text{x}}^{\rho_1}} 
\newcommand{\vxroo}{v_{\text{x}}^{\rho_1}} 
\newcommand{\hvecroo}{\hat{v}_{\text{ec}}^{\rho_1}} 
\newcommand{\hvec}{\hat{v}_{\text{ec}}} 
\newcommand{\vecroo}{v_{\text{ec}}^{\rho_1}} 
\newcommand{\tPsi}{\tilde{\Psi}}
\newcommand{\eps}{\epsilon} 
\newcommand{\hF}{\hat{F}} 
\newcommand{\cF}{{\cal F}} 
\newcommand{\hcF}{{\cal \hat{F}}} 
\newcommand{\hv}{\hat{v}} 
\newcommand{\hw}{\hat{w}} 
\newcommand{\nn}{\nonumber} 
\newcommand{\rco}{r_{12}^{-1}} 
\newcommand{\Eco}{E_{\text{co}}} 
\newcommand{\Ex}{E_{\text{x}}} 
\newcommand{\tEco}{\tilde{E}_{\text{co}}} 
\newcommand{\cEco}{{\cal E}_{\text{co}}} 
\newcommand{\vep}{\varepsilon} 

\title{State Specific Kohn--Sham Density Functional Theory}

\author{James P. Finley}

\affiliation{
Department of Physical Sciences,
Eastern New~Mexico University,
Station \#33, Portales, NM 88130}
\affiliation{
Department of Applied Chemistry,
Graduate School of Engineering,
The University of Tokyo\\
7-3-1 Hongo, Bunkyo-ku,
Tokyo, 113-8656 Japan}
\email{james.finley@enmu.edu}
\date{\today}

\begin{abstract}

A generalization of the Kohn--Sham approach is derived where the
correlation-energy functional depends on the one-particle density matrix of
noninteracting states and on the external potential from the interacting
target-state. The one-particle equations contain the exact exchange potential, a
nonlocal correlation potential, and an additional operator involving the
correlation density. The electronic-energy functional has multiple solutions: Any
one-particle density matrix delivering the target-state density yields a
solution. In order to obtain the Kohn--Sham solution, the nonlocal operators are
converted into local ones using an approach developed by Sala and G\"orling.
Since the exact exchange-potential is used, and the $N$--representability problem
does not arise---in contrast to the Kohn--Sham approach---errors from Coulomb
self-interactions do not occur, nor the need to introduce functionals defined by a
constraint search.  Furthermore, the approach does not use the Hohenberg-Kohn
theorem. A density functional formalism is also derived that assumes that the
one-particle density matrices of interest have $v$-representable
(non-interacting) densities and that these density matrices can be written as
an explicit functional of the electron density. For simplicity, we only consider
noninteracting closed-shell states and target states that are nondegenerate,
singlet ground-states.

\end{abstract}

\maketitle

\section{Introduction}

The Kohn-Sham version of density functional theory plays a major role in both
quantum chemistry and condensed matter physics
\cite{Dreizler:90,Parr:89,Springborg:97,Ellis:95,Gross:94,Seminario:95,Handy:97}.
The local density approximation \cite{Kohn:65} has been widely used for the solid
state. While for molecules, by far, the most successful functional, a hybrid one
\cite{Becke:93,Burke:97,Perdew:96,Ernzerhof:96}, is known as B3LYP
\cite{Becke:93,Stephens:94}. 

The Kohn--Sham approach, however, does have well known shortcomings. For example,
a constraint search definition \cite{Percus:78,Levy:79,Levy:82,Levy:85} is
required to treat the $v$--representability problem that arises in the original
Kohn--Sham method \cite{Kohn:65}.  Unfortunately, this formal definition is
difficult to consider when deriving approximate functionals. Furthermore, in
contrast to wave function based methods, the exchange-correlation functional is an
unknown, implicit functional, and there is no systematic method to improve
approximations. In addition, there are well known errors arising from Coulomb
self-interactions that appears when using approximate functionals
\cite{Parr:89,Dreizler:90,Koch:00}. Also, the most widely used approximate
functional for molecular systems, the B3LYP functional, includes a component of
the exact exchange-potential, even though the Kohn--Sham approach requires the
noninteracting state to come from a local potential. The optimized potential
method \cite{Fiolhais:03,Sharp:53,Talman:76,Li:93,Shaginyan,Goorling:94,Grabo:00}
is an approach to convert a nonlocal operator into a local potential.
Unfortunately, this method leads to potentials that are not invariant to a unitary
transformation of orbitals and depend explicitly on the individual orbitals and
orbital energies.

The formalism presented below uses an electronic-energy functional containing a
correlation energy functional $\cEco$ that depends on the external potential $v$
and on the one-particle density matrix $\roo$ of determinantal states. Since the
$v$--representability problem does not appear, a constrain search definition is
not needed. Also, since the approach uses the exact exchange-potential, errors
from Coulomb self-interactions do not occur.  The energy functionals, however,
contains multiple solutions, since any one-particle density matrix $\roo$
delivering the density from the interacting state yields a solution. In order to
obtain the Kohn--Sham solution, the nonlocal operators are converted into local
ones using an approach developed by Sala and G\"orling \cite{Sala:01}. In contrast
to the optimized potential method
\cite{Fiolhais:03,Sharp:53,Talman:76,Li:93,Shaginyan,Goorling:94,Grabo:00}, the
energy functionals and local potentials are invariant to a unitary transformation
of orbitals and do not depend on the individual orbital or the orbital energies.
A density functional formalism is also derived that assumes that the one-particle
density matrices of interest have $v$-representable (non-interacting) densities
and that these density matrices can be written as an explicit functional of the
electron density.

Previously we have shown that the correlation energy from many body perturbation
theory \cite{Lindgren:86,Harris:92,Raimes:72} can be written as an explicit
functional of $v$ and $\roo$ \cite{Finley:bdmt.arxiv}. In a similar manner, but
using less restrictive energy denominators, the correlation energy functionals
presented below can be shown to be an explicit functional of $v$ and $\roo$
\cite{Finley:tobe}. Hence, in contrast to the Kohn--Sham method, it maybe possible
to derive approximate functionals that can be improved in a systematic manner.
For simplicity, we only consider noninteracting closed-shell states and target
states that are nondegenerate, singlet ground-states.

\section{The energy functionals and trial wave functions}

Our interest is in finding the ground-state eigenvalue of the Hamiltonian operator,
\begin{eqnarray} \zlabel{8793}
\hH_{Nv} = \hT + \hVee + \hV_v,
\end{eqnarray}
where 
\begin{eqnarray}
\hT   &=& \sum_{i}^{N}(\ke{i}), \\
\hVee &=& \fc12 \sum_{i\ne j}^{N}\rc{ij}, \\ 
\hV_v &=& \sum_{i}^{N}v(i),
\end{eqnarray}
and $v$ is the external potential; $N$ is the number of electrons.  Since the
Hamiltonian $\hH_{Nv}$ is determined by $N$ and $v$, so are the ground state
wave functions $|\Psi_{Nv}\ran$ that satisfy the Schr\"odinger equation:
\begin{eqnarray}\zlabel{4729}
\hH_{Nv} |\Psi_{Nv}\ran = \cE_{Nv}|\Psi_{Nv}\ran,
\end{eqnarray}
where, for simplicity, we only consider wave functions that are nondegenerate,
singlet ground-states.

Using a second quantization approach, our spin-free Hamiltonian does not depend on
$N$, and it can be expressed by
\begin{eqnarray} \zlabel{0115}
\hH_v = 
\sum_{ij}(i|(\ke{})|j) \hE_{ij} + \sum_{ij}(i|v|j) \hE_{ij} + 
\fc12 \sum_{ijkl} (ij|kl) \hE_{ijkl},
\end{eqnarray}
where the symmetry-adapted excitation operators are given by
\begin{eqnarray} \zlabel{8732}
\hE_{ij}&=&\sum_{\si} a_{i\si}^\dg a_{j\si}, \\
\zlabel{7282}
\hE_{ijkl}&=&
\sum_{\si\lambda}
a_{i\si}^\dg a_{k\lambda}^\dg a_{l\lambda} a_{j\si},
\end{eqnarray}
and the one- and two electrons integrals are spin-free integrals written in
chemist's notation \cite{Szabo:82} using a spatial orbital set, say $\{\chi\}$;
this set has the following form:
\begin{eqnarray}\zlabel{6752}
\psi_{j\si}(\mx)= \chi_j(\mr)\si(\om); \;\; \si=\al,\be,
\end{eqnarray}
where the spatial and spin coordinates, $\mr$ and $\om$, are denoted collectively
by $\mx$.

Wave function-based methods including perturbation theory, configuration
interaction, and coupled cluster theory, use one or more reference states to
express $\Psi$ and $\cE$. For closed-shell ground-state wave functions, a single
determinant can be used, where closed-shell determinantal, or noninteracting,
states can be constructed from a set of doubly occupied spatial-orbitals; these
occupied orbitals also determine the spin-less one-particle density-matrix of the
noninteracting state, given by \cite{Parr:89,McWeeny:60}
\begin{eqnarray} \zlabel{7298}
\roo(\mr_1,\mr_2) = 
2 \sum_{w\in\{\chi_o\}}  \chi_{w}(\mr_1) \chi_{w}^*(\mr_2),
\end{eqnarray}
where the sum is over the occupied orbitals; this set of orbitals is denoted by
$\{\chi_o\}$.  

For later use, we also mention that for a complete basis set we have
\begin{eqnarray} \zlabel{9228}
2\dt(\mr_1-\mr_2) = \roo(\mr_1,\mr_2) + \kp_{\rho_1}(\mr_1,\mr_2),
\end{eqnarray}
where $\kp_{\rho_1}$ is determined by the excited orbitals,
\begin{eqnarray} 
\kp_{\rho_1}(\mr_1,\mr_2) =
2 \sum_{r\in\{\chi_u\}}  \chi_{r}(\mr_1) \chi_{r}^*(\mr_2),
\end{eqnarray}
and $\{\chi_u\}$ denotes the set of orbitals orthogonal to the occupied set
$\{\chi_o\}$.  The operator form of Eq.~(\ref{9228}) is
\begin{eqnarray} \zlabel{7376}
\mbox{\small $2$} \mbox{\small $\hat{I}$} = \hroo + \hat{\kp}_{\rho_1},
\end{eqnarray}
where $\mbox{\small $\hat{I}$}$ is the identity operator; so, the kernels of the
three operators within Eq.~(\ref{7376}) are given by the corresponding terms
within Eq.~(\ref{9228}).

It is well known that there is a one-to-one mapping between determinantal states
and their one-particle density matrices \cite{Parr:89,Blaizot:86}, say $\gm$,
where for a closed-shell state described by the orbitals given by
Eq.~(\ref{6752}), we have \cite{Dirac:30,Dirac:31,Lowdin:55a,Lowdin:55b}
\begin{eqnarray} \zlabel{5721}
\gm(\mx_1,\mx_2) = 
\sum_{w\in\{\chi_o\}}  
\sum_\si \chi_{w}(\mr_1) \chi_{w}^*(\mr_2) \si(\om_1) \si^*(\om_2),
\end{eqnarray}
and by using Eq.~(\ref{7298}), we obtain
\begin{eqnarray} \zlabel{9783}
\gm(\mx_1,\mx_2) = \fc12 \roo(\mr_1,\mr_2) \dt_{\om_1\om_2}.
\end{eqnarray}
Since our closed-shell determinantal states are determined by $\roo$, we denote
these kets by~$|\roo\ran$.

According to the Hohenberg-Kohn theorem \cite{Hohenberg:64a,Parr:89,Dreizler:90},
the external potential $v$ is determined by the density, and the density also
determines $N$.  So, in principle, we can replace the variables $N$ and $v$ by the
electronic density $n$ and, at least for nondegenerate ground-states, write
\begin{eqnarray} \zlabel{0283}
\hH_{v} |\Psi_{n}\ran = \cE_{n}|\Psi_{n}\ran; \;\; n \longrightarrow N,v,
\end{eqnarray}
where these functions serve as density-dependent trial-wave functions for the
Kohn-Sham approach. Notice we have omitted the $N$ subscript on the Hamiltonian
operator, since $\hH_{v}$ is independent of $N$ when this operator is expressed
in second quantization.

As an alternative to a density-dependent wave function, we consider trial
wave functions, say $|\tPsi_{v\roo}\ran$, that are determined by the one-body
external potential $v$ and, in addition, by the spin-less one-particle
density-matrix $\roo$ of a noninteracting state, and, as mentioned previously,
these  noninteracting states are denoted by $|\roo\ran$.

By definition, our trial wave function $|\tPsi_{v\roo}\ran$ yields the exact
ground-state wave function $|\Psi_{n}\ran$ when the noninteracting density
$\rho_s$, i.e., the density of $|\roo\ran$, equals the exact density $n$ of the
interacting state $|\Psi_{n}\ran$, where $n$ also determines the $v$ and $N$. This
state of affairs can be represented by the following:
\begin{eqnarray} \zlabel{7220}
|\tPsi_{v\roo}\ran = |\Psi_{n}\ran; 
\;\; \roo \longrightarrow  \rho_s = n,\;\; n \longrightarrow N,v. 
\end{eqnarray}
In other words, $\roo$ determines $\rho_s$, and when $\rho_s = n$, 
$|\tPsi_{v\roo}\ran$ yields $|\Psi_{n}\ran$. Letting $\vroo$ denote the
one-particle density matrix of interest, we can write
\begin{eqnarray} \zlabel{3810}
|\tPsi_{v\vroo}\ran = |\Psi_{n}\ran; 
\;\; \vroo \longrightarrow  n,\;\; n \longrightarrow N,v. 
\end{eqnarray}

For later use, we also mention that the density $n$ of an interacting state can be
partitioned as
\begin{eqnarray} \zlabel{4820}
n = \rho_s + \rho_c, 
\end{eqnarray} 
where the correlation density is given by
\begin{eqnarray} \zlabel{8291}
\rho_c(\mr) =  \fc{\lan \Psi_{n}|\hGm(\mr)|\Psi_{n}\ran}{\lan \Psi_{n}|\Psi_{n}\ran}
- \rho_s(\mr), 
\end{eqnarray} 
and $\hGm$ is the density operator, given by Eq.~(\ref{6211}).
 
Using our trial wave function, we introduce a variational energy functional:
\begin{eqnarray} \zlabel{3819}
E_v[\roo] = 
\fc{\lan \tPsi_{v\roo}|\hH_{v}|\tPsi_{v\roo}\ran}
   {\lan \tPsi_{v\roo}|\tPsi_{v\roo}\ran}.
\end{eqnarray}

Our trial wave functions $|\tPsi_{v\roo}\ran$ and energy functionals $E_v[\roo]$
are assumed to be explicit functionals of $\roo$ and $v$. However, two different
one-particle density matrices, say $\roo$ and $\roo^{\pr}$, that yield the same
density $\rho_s$, i.e., $\roo\longrightarrow \rho_s$ and $\roo^\pr \longrightarrow
\rho_s$, yield the same $|\tPsi_{v\roo}\ran$ and $E_v[\roo]$, so these functions
are implicit functionals of $\rho_s$, and, therefore, we can write
$|\tPsi_{v\rho_s}\ran$ and $E_v[\rho_s]$. However, we will continue to consider
them as functionals of their explicit variable $\roo$.

Using Eqs.~(\ref{0283}) and (\ref{3810}), we observe that our energy functional
$E_v$, given by Eq.~(\ref{3819}), delivers the exact energy $\cE_{n}$ when the
one-particle density matrix determines the exact density~$n$:
\begin{eqnarray} \zlabel{8273}
E_v[\vroo] = \cE_{n}, \;\; \vroo \longrightarrow n, 
\;\; n \longrightarrow N,v,
\end{eqnarray}
and for an arbitrary density we get
\begin{eqnarray}
E_v[\roo] \ge \cE_{n}, 
\;\; \roo \longrightarrow  \rho_s \longrightarrow N,
\end{eqnarray}
where the density $\rho_s$ from the noninteracting state $|\roo\ran$ is not
necessarily $v$-representable.

\section{Trial Hamiltonians}

Our trial wave function is a ground-state eigenfunction of a Hamiltonian operator
that depend explicitly on the one-particle density of a noninteracting state:
\begin{eqnarray} \zlabel{4273}
\hH_{v\roo}|\tPsi_{v\roo}\ran = \tilde{\cE}_{v\roo}|\tPsi_{v\roo}\ran.
\end{eqnarray}
As in our trial wave functions $|\tPsi_{v\roo}\ran$ and energy functionals
$E_v[\roo]$, the trial Hamiltonians $\hH_{v\roo}$ are explicit functionals of
$\roo$, but implicit functionals of $\rho_s$. So two trial Hamiltonians, say
$\hH_{v\roo}$ and $\hH_{v\roo^\pr}$, are equal if both $\roo$ and $\roo^\pr$ yield
the same density, i.e., $\roo,\roo^\pr \rightarrow \rho_s$.

Our trial Hamiltonians must be chosen so that Eq.~(\ref{3810}) is satisfied,
indicating the following identity:
\begin{eqnarray} \zlabel{4863}
\hH_{v\vroo} = \hH_{v},
\;\; \vroo \longrightarrow  n,\;\; n \longrightarrow N,v. 
\end{eqnarray}

There are many ways to obtain a trial Hamiltonian that satisfies
Eq.~(\ref{4863}). Consider the following trial Hamiltonian obtained by adding a
term to the Hamiltonian:
\begin{eqnarray} \zlabel{5289}
\hH_{v\roo} = \hH_v + \lambda\int d\mr\, g_{\rho_c}(\mr) 
\lt(\hGm(\mr) - \rho_s(\mr)\rt),
\;\; \roo \longrightarrow \rho_s,
\end{eqnarray}
where $\hGm(\mr)$ is the density operator, given by Eq.~(\ref{6211}); $(\hGm(\mr)
- \rho_s(\mr))$ is the one-body portion of $\hGm(\mr)$ when this operator is
written in normal-ordered form \cite{Cizek:66,Cizek:69,Lindgren:86,Paldus:75},
given by Eq.~(\ref{2623}).  Furthermore, $\lambda$ is an arbitrary constant, and
the functional $g$ is also arbitrary, except that it vanishes when the correlation
density $\rho_c$ vanishes
\begin{eqnarray} 
\lim_{\rho_c\to 0} g_{\rho_c}(\mr) = 0,
\end{eqnarray}
where $\rho_c$ is defined by Eqs.~(\ref{4820}) and (\ref{8291}).

Since $(\hGm(\mr) - \rho_s(\mr))$ is normal-ordered, we have
\begin{eqnarray} 
\lan \roo |\lt(\hGm(\mr) - \rho_s(\mr)\rt)|\roo \ran = 0.
\end{eqnarray}
Therefore, the added term appearing in Eq.~(\ref{5289}) can be considered a sort
of correlation term, since it does not contribute in first order. Hence, we have
\begin{eqnarray} 
\lan \roo |\hH_{v\roo}|\roo \ran =  \lan \roo|\hH_v|\roo\ran. 
\end{eqnarray}

One possible choice for $g_{\rho_c}$, and presented in Appendix~\ref{2318}, is given by
\begin{eqnarray} \zlabel{8271}
g_{\rho_c}(\mr_1) = \int d\mr_2\, \rc{12}\, \rho_c(\mr_2).
\end{eqnarray}

\section{A generalization of the Kohn-Sham formalism}

We now obtain a generalization of the Kohn-Sham formalism. Substituting
Eq.~(\ref{8793}) into Eq.~(\ref{3819}) gives
\begin{eqnarray} \zlabel{5428}
E_v[\roo] = 
  \fc{\lan \tPsi_{v\roo}|\hT|\tPsi_{v\roo}\ran}
   {\lan \tPsi_{v\roo}|\tPsi_{v\roo}\ran}
+ \fc{\lan \tPsi_{v\roo}|\hVee|\tPsi_{v\roo}\ran}
   {\lan \tPsi_{v\roo}|\tPsi_{v\roo}\ran}
+ \int d\mr\, v(\mr) \rho_s(\mr)
+ \int d\mr\, v(\mr) \tilde{\rho}_c(\mr),
\end{eqnarray}
where $\tilde{\rho}_c$ is the correlation density of the trial wave function, i.e,
as in Eq.~(\ref{8291}), we have
\begin{eqnarray} 
\tilde{\rho}_c(\mr) =  
\fc{\lan \tPsi_{v\roo}|\hGm(\mr)|\tPsi_{v\roo}\ran}
   {\lan \tPsi_{v\roo}|\tPsi_{v\roo}\ran} - \rho_s(\mr)
= \tilde{n} - \rho_s(\mr), 
\;\; \tPsi_{v\roo} \longrightarrow \tilde{n}, \roo \longrightarrow \rho_s,
\end{eqnarray} 
and $\tilde{n}$ is the density of $\tPsi_{v\roo}$.

Through the first-order, the kinetic energy and electron-electron repulsion energy are
given, respectively, by
\begin{eqnarray} 
\lan \roo|\hT|\roo \ran&=& \int d\mr_1\, \lt[\ke{1}\roo(\mr_1,\mr_2)\rt]_{\mr_2=\mr_1}, \\
\lan \roo|\hVee|\roo \ran &=& E_J[\rho_s] + \Ex[\roo],
\end{eqnarray}
where the Coulomb and exchange energies are
\begin{eqnarray} 
E_J[\rho_s] &=& \fc12 \int \int \rco d\mr_1 d\mr_2 \rho(\mr_1)\, \rho(\mr_2), \\ 
-\Ex[\roo] &=& \fc14 \int \int \rco d\mr_1 d\mr_2 \roo(\mr_1,\mr_2)\, \roo(\mr_2,\mr_1).
\end{eqnarray}
Adding and subtracting $\lan \roo|\hT|\roo\ran$ and $\lan \roo|\hVee|\roo \ran$,
Eq.~(\ref{5428}) can be written as
\begin{eqnarray} \label{1272}
E_v[\roo] = \int d\mr_1\, \lt[\ke{1}\roo(\mr_1,\mr_2)\rt]_{\mr_2=\mr_1}
+ \int d\mr\, v(\mr) \rho_s(\mr) 
\hs{30}
\nn \\ \mbox{}
\hs{10}
+ E_J[\rho_s] + \Ex[\roo] + \Eco[\roo,v]
+ \int d\mr\, v(\mr) \tilde{\rho}_c(\mr),
\end{eqnarray}
where the correlation-energy functional is given by
\begin{eqnarray} \zlabel{1928}
\Eco[\roo,v] =  
\fc{\lan \tPsi_{v\roo}|\hT|\tPsi_{v\roo}\ran}
   {\lan \tPsi_{v\roo}|\tPsi_{v\roo}\ran}
-  \lan \roo|\hT|\roo \ran
+ \fc{\lan \tPsi_{v\roo}|\hVee|\tPsi_{v\roo}\ran}
   {\lan \tPsi_{v\roo}|\tPsi_{v\roo}\ran}
- \lan \roo|\hVee|\roo \ran.
\end{eqnarray}
Recognizing the first four terms from Eq.~(\ref{1272}) as the energy through the
first order, $\cE_1$, we can write
\begin{eqnarray} \zlabel{8281}
E_v[\roo] = \cE_1[\roo,v] + \Eco[\roo,v] + \int d\mr\, v(\mr) \tilde{\rho}_c(\mr),
\end{eqnarray}
where
\begin{eqnarray} \label{4217}
\cE_1[\roo,v] = \lan \roo|H_{v}|\roo \ran = 
\int d\mr_1\, \lt[\ke{1}\roo(\mr_1,\mr_2)\rt]_{\mr_2=\mr_1} 
\hs{35}\\
\nn \hs{0.5}  \mbox{}
+   \int d\mr_1\, v(\mr_1)\rho(\mr_1) 
+   \fc12 \int \int d\mr_1 d\mr_2 \rco \rho(\mr_1)\, \rho(\mr_2) 
-   \fc14 \int \int d\mr_1 d\mr_2 \rco \roo(\mr_1,\mr_2)\, \roo(\mr_2,\mr_1).
\end{eqnarray}

Now consider the correlation energy that is obtained by wave function
methods. Using the notation from Eq.~(\ref{4729}), and a reference state
$|\roo\ran$, the correlation energy is given by
\begin{eqnarray}
\cEco[\roo,v] =
\fc{\lan \Psi_{Nv}|\hH_{v}|\Psi_{Nv}\ran}{\lan \Psi_{Nv} |\Psi_{Nv}\ran} 
- \cE_1[\roo,v],
\end{eqnarray}
where previously we have shown that $\cEco$ can be written as an explicit
functional of $v$ and $\roo$ \cite{Finley:bdmt.arxiv}. In a similar manner, but
using less restrictive energy denominators, our correlation energy functional
$\Eco$, given by Eq.~(\ref{1928}), can be shown to be an explicit functional of
$v$ and $\roo$ \cite{Finley:tobe}. Therefore, by requiring the last term within
Eq.~(\ref{8281}) to be an explicit functional of $v$ and $\roo$, $E_v$ can also be
written as an explicit functional of $v$ and $\roo$ \cite{Finley:tobe}.

We now focus our attention on minimizing the energy functional $E_v$, subject to
the constraint that the spin-less one-particle density-matrix $\roo$ comes from a
closed-shell single-determinantal state. For the more general case of a
determinantal state, say $|\gm\ran$, with the (spin-dependent) one-particle
density matrix $\gm$, as in Eq.~(\ref{5721}), the two necessary conditions for
$\gm$ to satisfy are given by the following \cite{Blaizot:86,Parr:89}:
\begin{eqnarray} 
\int \int \gm(\mx_3,\mx_4) \dt(\mx_3-\mx_4)\, d\mx_3 d\mx_4 = N, \\
\int \gm(\mx_3,\mx_5) \gm(\mx_5,\mx_4)\, d\mx_5 = \gm(\mx_3,\mx_4),
\end{eqnarray}
where the first relation indicates that the electron density yields the number of
electrons $N$; the second relation indicates that $\gm$ is indempotent. For our
special closed-shell case, we substitute Eq.~(\ref{9783}) into the above
constrains, yielding the following conditions:
\begin{eqnarray} \zlabel{5271}
\int \int \roo(\mr_3,\mr_4) \dt(\mr_3-\mr_4)\, d\mr_3 d\mr_4 = N, \\
\zlabel{5292}
\int \roo(\mr_3,\mr_5) \roo(\mr_5,\mr_4)\, d\mr_5 = 2\roo(\mr_3,\mr_4).
\end{eqnarray}

It is well know that the functional derivative of $\cE_1$ with respect to the
$\gm$ yields the kernel of the Fock operator \cite{Parr:89}. For the closed-shell case,
we have
\begin{eqnarray} \label{4281}
F(\mr_1,\mr_2) =
\fc{\dt \cE_1[\roo,v]}{\dt \roo(\mr_2,\mr_1)},
\end{eqnarray}
where, using Eq.~(\ref{4217}), the Fock kernel is given by
\begin{eqnarray} 
F_\roo(\mr_1,\mr_2) =
\dt(\mr_1-\mr_2)
\lt( 
 \ke{2} + v(\mr_2) + \int d\mr_3 \rc{23} \rho(\mr_3) 
\rt)
+ \vxroo(\mr_1,\mr_2),
\end{eqnarray}
and the exchange operator, say $\hvxroo$, has the following kernel:
\begin{eqnarray} 
\vxroo(\mr_1,\mr_2) = - \fc12 \rco \roo(\mr_1,\mr_2).
\end{eqnarray}
By generalizing Eq.~(\ref{4281}), we define a generalized, or exact, Fock
operator $\hcF$, where the kernel of this operator is
\begin{eqnarray} 
\cF_{\roo} (\mr_1,\mr_2) =
\fc{\dt E_v[\roo]}{\dt \roo(\mr_2,\mr_1)} = F_\roo(\mr_1,\mr_2) 
+ \vcoroo(\mr_1,\mr_2) + \vecroo(\mr_1,\mr_2), 
\end{eqnarray}
and the correlation operator $\hvcoroo$ and external-correlation operator
$\hvecroo$ are defined by their kernels:
\begin{eqnarray} 
\vcoroo(\mr_1,\mr_2)&=& \fc{\dt \Eco[\roo,v]}{\dt \roo(\mr_2,\mr_1)}, \\
\vecroo(\mr_1,\mr_2)&=& 
\fc{\dt \lt(\int d\mr_3\, v(\mr_3) \tilde{\rho}_c(\mr_3)\rt)}{\dt \roo(\mr_2,\mr_1)}.
\end{eqnarray}
 
Minimizing the functional $E_v$, given by Eq.~(\ref{8281}), subject to the
constraints given by Eqs.~(\ref{5271}) and (\ref{5292}), is very similar to the
corresponding Hartree--Fock derivation \cite{Parr:89} and the derivation for
reference-state one-particle density matrix theory
\cite{Finley:bdmt,Finley:bdmt.arxiv,Finley:bdft}. The only difference being that
the spin variable has been eliminated, and we have a factor of two appearing in
Eq.~(\ref{5292}). Therefore, we only state the main results, i.e., this
minimization yields the exact electronic energy $\cE_{n}$ for the interacting
state, as given by Eq.~(\ref{8273}), where the one-particle density-matrix $\vroo$
satisfies the following conditions:
\begin{eqnarray} \zlabel{8591}
\hkp_{\varrho_1} \hcF_{\vroo} \hat{\varrho}_1 &=&0, \\
\zlabel{8592}
\hat{\varrho}_1 \hcF_{\vroo} \hkp_{\varrho_1} &=&0,
\end{eqnarray}
and the kernels of the operators $\hroo$ and $\hat{\kp}_{\rho_1}$ are given by the
terms on the right side of Eq.~(\ref{9228}); also, as mentioned previously,
$\vroo$ yields the exact density $n$ of the interacting state~$\Psi_{n}$.  Using
Eqs.~(\ref{8591}) and (\ref{8592}), it is readily shown that $\hcF_{\vroo}$ and
$\hvroo$ commute:
\begin{eqnarray} 
\lt[\hcF_{\vroo},\hat{\varrho}_1\rt] =0,
\end{eqnarray}
and the occupied orbitals satisfy a generalized Hartree--Fock equation:
\begin{eqnarray} 
\hcF_{\vroo} \chi_{w} =   \sum_{x\in\vroo} \vep_{xw} \chi_{x},  
\end{eqnarray}
where the notation $x\in\vroo$ indicates a summation over the occupied orbitals
from the determinantal state $|\vroo\ran$; $\chi_{w}$ is also an occupied orbital
from $|\vroo\ran$. Furthermore, we can choose orbitals that diagonalize the matrix
$\vep_{xw}$, yielding exact, canonical Hartree--Fock equations:
\begin{eqnarray} \zlabel{0528}
\lt(\ke{} + v + v_j^{n} + \hvx^{\vroo} + \hvco^{\vroo} + \hvec^{\vroo} \rt)
\chi_{w} =   \vep_{w} \chi_{w}, 
\;\; \chi_w\in\vroo,
\end{eqnarray}
where the Coulomb operator is defined by
\begin{eqnarray} 
v_j^{\rho}(\mr_1)  \chi(\mr_1) =
\int d\mr_2 \rc{12} \rho(\mr_2)  \chi(\mr_1),
\end{eqnarray}
and we have
\begin{eqnarray} 
\vroo(\mr,\mr) = n(\mr).
\end{eqnarray}
Equation~(\ref{0528}) is also satisfied by the canonical excited orbitals.

For later use, we also mention that the determinantal states $|\vroo\ran$ satisfy
the following noninteracting Schr\"odinger equation:
\begin{eqnarray} \zlabel{7829}
\sum_{i=1}^N \hcF_{\vroo}(\mr_i) |\vroo\ran =   2\lt(\sum_w\vep_{w}\rt) |\vroo\ran.
\end{eqnarray}

Appendix \ref{3281} presents an alternative way of partitioning the energy
functional that differs from Eq.~(\ref{8281}).

\section{Conversion of the nonlocal potential into a local one}

As mentioned previously, our energy functionals $E_v$ are implicit functionals of
the noninteracting density $\rho_s$. Hence, any one-particle density-matrix that
yields the interacting density minimizes our energy functional, i.e.,  we have
\begin{eqnarray} \zlabel{5761}
\cE_{n} = E_v[\vroo] = E_v[\vroo^{\pr}] = E_v[\vroo^{\pr\pr}] \cdots,
\end{eqnarray}
where
\begin{eqnarray} \zlabel{5789}
n(\mr) = \vroo(\mr,\mr) = \vroo^{\pr}(\mr,\mr) = \vroo^{\pr\pr}(\mr,\mr) \cdots,
\end{eqnarray}
and there are other solutions besides Eq.~(\ref{0528}), e.g,
\begin{eqnarray} \zlabel{1329}
\hcF_{\vroo^{\pr}} \chi_{w} =
\lt(\ke{} + v + 
v_j^{n} + \hat{w}_{\vroo^{\pr}}
\rt)
\chi_{w} =   \vep_{w}^{\pr} \chi_{w}, 
\;\; \chi_w\in\vroo^\pr,
\end{eqnarray}
where the nonlocal potential $\hat{w}_{\roo}$ is given by
\begin{eqnarray} 
\hat{w}_{\roo} = \hvx^{\roo} + \hvco^{\roo} + \hvec^{\roo}.
\end{eqnarray}

Assuming $n$ is a noninteracting $v$-representable density, there exist a
noninteracting state, say $|\varphi_1\ran$, that has $n$ as its density:
\begin{eqnarray} \zlabel{5772}
n(\mr)=\varphi_1(\mr,\mr),
\end{eqnarray}
and this determinant---assuming it is a closed-shell determinant---is the
ground-state solution of the following noninteracting Schr\"odinger equation:
\begin{eqnarray} \zlabel{7811}
\sum_{i=1}^N \hat{f}(\mr_i) |\varphi_1\ran =   
2\lt(\mbox{\small $\displaystyle \sum_w$}\eps_{w}\rt) |\varphi_1\ran,
\end{eqnarray}
where
\begin{eqnarray} 
\hat{f} = \ke{} + v_s,
\end{eqnarray}
and $v_s$ is a local potential. Therefore, the canonical occupied orbitals from
$|\varphi_1\ran$ satisfy the following one-particle Schr\"odinger equation:
\begin{eqnarray} \zlabel{3292}
\hat{f} \phi_{w} =
\lt(\ke{} + v + v_j^{n} + \vxc \rt)
\phi_{w} =   \eps_{w} \phi_{w}, 
\;\; \phi_w\in\varphi_1,
\end{eqnarray}
where with no loss of generality, we have required $v_s$ to be defined by
\begin{eqnarray} 
v_s = v + v_j^{n} + \vxc.
\end{eqnarray}

By definition, or using Eqs.~(\ref{5761}), (\ref{5789}), and (\ref{5772}),
$\varphi_1$ is a one-particle density matrix that minimizes our energy functional:
\begin{eqnarray} 
\cE_{n} = E_v[\varphi_1],
\end{eqnarray}
and, therefore, $\varphi_1$ also satisfies Eq.~(\ref{7829}):
\begin{eqnarray} \zlabel{7844}
\sum_{i=1}^N \hcF_{\varphi_1}(\mr_i) |\varphi_1\ran =   2\lt(\sum_w\eps_{w}\rt) |\varphi_1\ran.
\end{eqnarray}
Hence, it follows from Eqs.~(\ref{7811}) and (\ref{7844}) that $|\varphi_1\ran$ is
an eigenstate of two different noninteracting Hamiltonians. By comparing
Eq.~(\ref{1329}) and (\ref{3292}) with $\vroo^{\pr}=\varphi_1$, we see that
the two operators, $\hcF_{\varphi_1}$ and $\hat{f}$, are identical, except that
$\hcF_{\varphi_1}$ contains the nonlocal operator $\hat{w}_{\varphi_1}$ and
$\hat{f}$ contains the local potential $\vxc$. Furthermore, the occupied orbitals
from Eq.~(\ref{1329}) and (\ref{3292}) with $\vroo^{\pr}=\varphi_1$ may differ by
a unitary transformation, but they yield the same one-particle density matrix:
\begin{eqnarray} 
\varphi_1(\mr_1,\mr_2) = 
2 \sum_{w\in\varphi_1}  \chi_{w}(\mr_1) \chi_{w}^*(\mr_2)
=
2 \sum_{w\in\varphi_1}  \phi_{w}(\mr_1) \phi_{w}^*(\mr_2).
\end{eqnarray}

Using the approach by Sala and G\"orling \cite{Sala:01}, and Eqs.~(\ref{7811}),
(\ref{7844}), (\ref{1329}) and (\ref{3292}), but permitting the orbitals to be
complex, it is readily demonstrated that $\vxc$ is given by
\begin{eqnarray} \label{2629}
\vxc(\mr) 
=  \fc{1}{2n(\mr)}
\int d\mr_1
\lt[2w(\mr_1,\mr) \varphi_1(\mr,\mr_1)
-   \varphi_1(\mr,\mr_1) \rt. \int d\mr_2 \, \varphi_1(\mr_2,\mr) w(\mr_1,\mr_2)
\hs{10}
\\ \nonumber \mbox{} 
\hs{50}
 \lt.
+ \varphi_1(\mr_1,\mr) \varphi_1(\mr,\mr_1) \vxc(\mr_1)\rt].
\end{eqnarray}
By substituting $\vxc$ repeatedly on the right side we can obtain an expansion for
$\vxc$:
\begin{eqnarray} \nn
\vxc(\mr) 
=  \fc{1}{2n(\mr)}
[2w(\mr_1,\mr) \varphi_1(\mr,\mr_1)
-   \varphi_1(\mr,\mr_1) \varphi_1(\mr_2,\mr) w(\mr_1,\mr_2) 
\hs{25}
\\ \nn \mbox{}
+ \varphi_1(\mr_1,\mr) \varphi_1(\mr,\mr_1) 
\fc{1}{n(\mr_1)}\{w(\mr_2,\mr_1) \varphi_1(\mr_1,\mr_2)
- \fc12\varphi_1(\mr_1,\mr_2)  \varphi_1(\mr_3,\mr_1) w(\mr_2,\mr_3)\}
\\ \label{1092} \mbox{}
+ \varphi_1(\mr_1,\mr) \varphi_1(\mr,\mr_1) 
\fc{1}{2n(\mr_1)} \varphi_1(\mr_2,\mr_1) \varphi_1(\mr_1,\mr_2) \fc{1}{n(\mr_2)}
w(\mr_3,\mr_2) \varphi_1(\mr_2,\mr_3) + \cdots 
], \hs{1}
\end{eqnarray}
where there are integrations over the dummy variables $\mr_1$, $\mr_2$ and
$\mr_3$. The leading term of Eq.~(\ref{1092}) is the Slater potential
\cite{Slater:51,Harbola:93,Hirata:01}; this term also appears within the
Krieger--Li--Iafrate (KLI) approximation of the optimized potential method
\cite{Fiolhais:03,Li:92,Li:93,Hirata:01}.

The orbitals $\phi_{w}$ satisfying Eq.~(\ref{3292}) are the Kohn--Sham orbitals
\cite{Kohn:65}; $|\varphi_1\ran$ is the Kohn--Sham noninteracting state. However,
$\hat{f}$ differs from the Kohn--Sham operator, since, in addition to depending
explicitly $\varphi_1$, instead of $n$, $\hat{f}$ depends explicitly on the
external potential $v$ from the interacting Hamiltonian $\hH_v$. Furthermore, the
external-correlation operator $\hvecroo$ does not appear in Kohn--Sham
formalism. In addition, unlike the original Kohn--Sham approach \cite{Kohn:65},
the $N$-representability problem does not arise, nor the need to introduce a
constraint-search definition \cite{Percus:78,Levy:79,Levy:82,Levy:85} to avoid
this problem.

In our derivation we have assumed that $|\varphi_1\ran$ is a ground state solution
of Eq.~(\ref{7811}). However, the results may also be valid if $|\varphi_1\ran$ is
an excited state solution, since the Sala and G\"orling approach may also be valid
in this case.

\section{Conversion of the one-particle density-matrix functionals into density functionals}

For noninteracting states, the wave function is determined by the one-particle
density matrix. For certain closed-shell determinantal states, we can write
$\roo[\rho_s]$, where this functional includes all densities that are
noninteracting $v$-representable, but it is also defined for all $N$-representable
densities. Using the constraint search approach
\cite{Percus:78,Levy:79,Levy:82,Levy:85}, for a given density, say $\rho^{\pr}$,
the functional $\roo[\rho^\pr]$ yields the one-particle density matrix that
minimizes the expectation value of the kinetic energy:
\begin{eqnarray} 
\text{Min}\put(-22,-7){\scriptsize $\roo \rightarrow \rho^\pr$}
\hspace{2ex} 
\lan \roo | \hT | \roo \ran =  
\lan \roo\mbox{\small $[\rho^\pr]$} | \hT | \roo\mbox{\small $[\rho^\pr]$} \ran,
\end{eqnarray}
where the search is over all determinantal states that have a density of
$\rho^\pr$.

Substituting $\roo[\rho]$ into $\Eco$ of Eq.~(\ref{1272}) gives
\begin{eqnarray} \label{5831}
E_v[\roo] = \int d\mr_1\, \lt[\ke{1}\roo(\mr_1,\mr_2)\rt]_{\mr_2=\mr_1}
+ \int d\mr\, v(\mr) \rho_s(\mr) 
\hs{30}
\nn \\ \mbox{}
+ E_J[\rho_s] + \Ex[\rho] + \Eco[\rho,v]
+ \int d\mr\, v(\mr) \tilde{\rho}_c(\mr),
\;\;  \rho \longrightarrow \roo,
\end{eqnarray}
where, using $\roo[\rho]$, the last term is also a functional of $v$ and
$\rho$. This equation differs from the Kohn--Shan density functional, since the
correlation-energy functional depends on the external potential $v$, and the last
term does not appear in the Kohn--Sham approach. However, mathematically speaking,
the minimization of Eq.~(\ref{5831}) follows the same procedure as in the
Kohn--Sham method, yielding
\begin{eqnarray} 
\hat{f} \phi_{w} =
\lt(\ke{} + v + v_j^{n} + v_{\text{x}}^n + \vco^n + v_{\text{ec}}^{n} \rt)
\phi_{w} =   \eps_{w} \phi_{w}, 
\;\; \phi_w\in\varphi_1,
\end{eqnarray}
where the local potentials are given by
\begin{eqnarray} 
v_{\text{x}}^{\rho}(\mr)&=& \fc{\dt \Ex[\rho,v]}{\dt \rho(\mr)}, \\
\vco^{\rho}(\mr)&=& \fc{\dt \Eco[\rho,v]}{\dt \rho(\mr)}, \\
v_{\text{ec}}^{\rho}(\mr)&=& 
\fc{\dt \lt(\int d\mr_1\, v(\mr_1) \tilde{\rho}_c(\mr_1)\rt)}{\dt \rho(\mr)}.
\end{eqnarray}

Assuming the density $n$ from the interacting state is noninteracting
$v$-representable, we have
\begin{eqnarray} 
E_v[n] = \cE_{n}, \;\; \mbox{$n$ is noninteracting $v$-representable}.
\end{eqnarray}
Note that Eq.~(\ref{5831}) is a valid energy functional only when the one-particle
density matrix that enters the first term is the same one generated by the
functional $\roo[\rho]$; this is the case, at least when $\rho$ is non-interacting
$v$-representable.

\appendix

\section{A possible choice for \mbox{\large $\lowercase{g}_{\rho_c}$} \zlabel{2318}}

The electron-electron repulsion operator is spin-free and can be written as
\begin{eqnarray}
\hVee = \fc12 \sum_{ij}(ij|\rc{12}|kl) \hE_{ijkl},
\end{eqnarray}
where the two-electron integral is written in chemist's notation \cite{Szabo:82}
and the two-electron spin-adapted excitation-operator is given by
Eq.~(\ref{7282}). This operator can also be written as
\begin{eqnarray} \zlabel{2915}
\hVee = \int\int d\mr_1 d\mr_2\, \rc{12}\, \hGm_2(\mr_2,\mr_1),
\end{eqnarray}
where the pair-function operator  is given by
\begin{eqnarray}\zlabel{8262}
\hGm_2(\mr_2,\mr_1) 
= \fc12 \sum_{ijkl} 
\chi_{j}(\mr_1)  \chi_{i}^*(\mr_1)
\chi_{l}(\mr_2) \chi_{k}^*(\mr_2)  
\hE_{ijkl},
\end{eqnarray}
and this operator yields the diagonal elements of the spinless two-particle
density matrix as the expectation value. Writing this operator in normal-ordered
form \cite{Cizek:66,Cizek:69,Lindgren:86,Paldus:75} with respect to the vacuum
state $|\roo\ran$, we have
\begin{eqnarray} \label{5282}
\hVee = 
  \int\int d\mr_1 d\mr_2\, \rc{12}\, \rho_2(\mr_2,\mr_1)_\roo 
+ \int\int d\mr_1 d\mr_2\, \rc{12}\, \rho_s(\mr_2)\hGm(\mr_1)_{\rho_s}
\hs{24}
\nonumber \\ \mbox{}
- \fc12 \int\int d\mr_1 d\mr_2\, \rc{12}\, \roo(\mr_2,\mr_1)\hGm(\mr_1,\mr_2)_{\roo}
+ \int\int d\mr_1 d\mr_2\, \rc{12}\, \hGm_2(\mr_2,\mr_1)_\roo,
\hs{2}
\end{eqnarray}
where, examining each term in turn, from the first term we have
\begin{eqnarray} \zlabel{4894}
\rho_2(\mr_2,\mr_1)_\roo = 
\fc12 \rho_s(\mr_2)\rho_s(\mr_1) 
-
\fc14 \roo(\mr_2,\mr_1) \roo(\mr_1,\mr_2),
\end{eqnarray}
and this function is the diagonal elements of the spinless second-order density
matrix of the determinantal state $|\roo\ran$. From the second term, we have
\begin{eqnarray} \zlabel{2623}
\hGm(\mr)_{\rho_s} = 
\sum_{ij}  \chi_{j}(\mr) \chi_{i}^*(\mr) \{\hE_{ij}\}_\roo,
\;\; \roo \longrightarrow \rho_s,
\end{eqnarray} 
and this operator is the one-body portion of the density operator, where the
density operator is given by
\begin{eqnarray} \zlabel{6211}
\hGm(\mr) = 
\sum_{ij}  \chi_{j}(\mr) \chi_{i}^*(\mr)\hE_{ij}.
\end{eqnarray} 
Note that we can write
\begin{eqnarray} \zlabel{3478}
\hGm(\mr)_{\rho_s} = \hGm(\mr) - \rho_s(\mr),
\end{eqnarray}
indicating that $\hGm(\mr)_{\rho_s}$ is determined by $\rho_s$ and not by $\roo$; two
different one-particle density matrices that yield the same density have the same
$\hGm(\mr)_{\rho_s}$.

Returning to Eq.~(\ref{5282}), from the third term we have
\begin{eqnarray} \zlabel{7392}
\hGm(\mr_1,\mr_2)_{\roo} = \sum_{ij} \chi_{j}(\mr_1) \chi_{i}^*(\mr_2)
\{\hE_{ij}\}_\roo,
\end{eqnarray}
and this operator is the one-body portion of the one-particle density-matrix
operator, given by
\begin{eqnarray}\zlabel{1828}
\hGm(\mr_1,\mr_2) = 
\sum_{ij}  \chi_{j}(\mr_1) \chi_{i}^*(\mr_2) \hE_{ij}
= \roo(\mr_1,\mr_2) +\hGm(\mr_1,\mr_2)_\roo.
\end{eqnarray}
And from the last term, we have
\begin{eqnarray}
\hGm_2(\mr_2,\mr_1)_\roo 
= \fc12 \sum_{ijkl} 
\chi_{j}(\mr_1)  \chi_{i}^*(\mr_1)
\chi_{l}(\mr_2) \chi_{k}^*(\mr_2) \{\hE_{ijkl}\}_\roo,
\end{eqnarray}
and this operator is the two-body portion of the pair-function operator,
Eq.~(\ref{8262}).

To obtain a slight modification of $\hVee$, we replace the determinantal state
density $\rho_s$, that appears in Eq.~(\ref{5282}), with the exact density $n$,
giving
\begin{eqnarray} 
\hVee^\roo = 
  \int\int d\mr_1 d\mr_2\, \rc{12}\, \rho_2(\mr_2,\mr_1)_\roo 
+ \int\int d\mr_1 d\mr_2\, \rc{12}\, n(\mr_2)\hGm(\mr_1)_{\rho_s}
\hs{24}
\nonumber \\ \mbox{}
- \fc12 \int\int d\mr_1 d\mr_2\, \rc{12}\, \roo(\mr_2,\mr_1)\hGm(\mr_1,\mr_2)_{\roo}
+ \int\int d\mr_1 d\mr_2\, \rc{12}\, \hGm_2(\mr_2,\mr_1)_\roo,
\hs{3}
\end{eqnarray}
and this operator can also be written as
\begin{eqnarray} 
\hVee^\roo = \hVee + 
\int\int d\mr_1 d\mr_2\, \rc{12}\, \rho_c(\mr_2)\lt(\hGm(\mr_1) - \rho_s(\mr_1)\rt),
\end{eqnarray}
Replacing $\hVee$ by $\hVee^\roo$ within the Hamiltonian operator, we have obtain
a trial Hamiltonian:
\begin{eqnarray} 
\hH_{v\roo} =  \hH_v 
+ \lambda \int\int d\mr_1 d\mr_2\, \rc{12}\, \rho_c(\mr_2)\lt(\hGm(\mr_1) - \rho_s(\mr_1)\rt),
\end{eqnarray}
where $\lambda$ is unity, but it can be permitted to be any constant
value. Comparing this equation with Eq.~(\ref{5289}) yields Eq.~(\ref{8271}).

\section{Energy Functional using Intermediate Normalization \zlabel{3281}}

Using Eq.~(\ref{5289}), our energy functional $E_v$, Eq.~(\ref{3819}), can be
also be written as
\begin{eqnarray} \zlabel{9026}
E_v[\roo] = 
  \fc{\lan \tPsi_{v\roo}|\hH_{v\roo}|\tPsi_{v\roo}\ran}
   {\lan \tPsi_{v\roo}|\tPsi_{v\roo}\ran}
- \lambda\int d\mr\, g_{\rho_c}(\mr)\lt(\hGm(\mr) - \rho_s(\mr)\rt).
\end{eqnarray}
By requiring our trial wave functions to satisfy intermediate normalization,
\begin{eqnarray}
\lan \roo |\tPsi_{v\roo}\ran = 1,
\end{eqnarray}
we have
\begin{eqnarray} 
E_v[\roo] = 
\lan \roo |\hH_{v\roo}|\tPsi_{v\roo}\ran
- \lambda\int d\mr\, g_{\rho_c}(\mr) \lt(\hGm(\mr) - \rho_s(\mr)\rt).
\end{eqnarray}
This form suggest the following partitioning:
\begin{eqnarray} 
E_v[\roo] = 
\cE_1[\roo,v] + \tEco[\roo,v]
- \lambda\int d\mr\, g_{\rho_c}(\mr) \lt(\hGm(\mr) - \rho_s(\mr)\rt),
\end{eqnarray}
where $\tEco$ is the correlation-energy (functional) of the trial wave function:
\begin{eqnarray} 
\tEco[\roo,v] = \lan \roo |\hH_{v\roo}|\tPsi_{v\roo}^Q\ran,
\end{eqnarray}
and the correlation function $\tPsi_{v\roo}^Q$ is defined by
\begin{eqnarray} 
|\tPsi_{v\roo}\ran = |\roo\ran + |\tPsi_{v\roo}^Q\ran.
\end{eqnarray}


\bibliography{ref}
\end{document}